\documentclass[aps,prd,preprint,superscriptaddress]{revtex4-1}
\usepackage[margin=1in]{geometry}
\usepackage{graphicx, subfig}

\begin{document}
\title{Quark Mass Matrices, textures and CKM precision measurements}
\author{Rohit Verma}
\email{rohitverma@live.com}
\affiliation{Chitkara University, Kalu Jhanda, Barotiwala, Himachal Pradesh 174103, India }
\date{\today}
\begin{abstract}

In the light of several recent analyses pointing towards texture 4-zero Fritzsch-like quark mass matrices as the only viable structures for quark mass matrices, this work adopts a model independent approach to reconstruct an alternate and simplified structure of texture specific quark mass matrices in a generalized 'u-diagonal' basis within the Standard Model framework using the Unitarity of CKM matrix and the observed hierarchies in quark mass spectra and mixing angles. It is observed that the measured 1$\sigma$ values of the three physical parameters namely $m_{\rm{u}}$, $m_{\rm{d}}$ and $s_{12}$ naturally lead to the vanishing  of (11) element in the down type quark mass matrix and that the single measurable CP violating phase $\delta_{13}$ in the CKM matrix is sufficient enough in $M^{\prime}_{\rm{d}}$ to explain the observed mixing pattern in a suitable basis. The leading order as well as exact analytic phenomenological solutions are addressed for the modest pattern of quark mass matrices derived from CKM matrix and precision measurements of mixing parameters.
\end{abstract}

\maketitle
\section{Introduction}
In the Standard Model (SM) framework \cite{Glashow:1961tr, Weinberg:1967tq, Salam:1968rm}, the couplings between the Higgs and quark fields are explained through the Lagrangian term
\begin{eqnarray}\label{Lagrangian}
-{\cal L}^{}_{\rm Y} = \overline{Q^{}_{\rm L}} Y^{}_{\rm u} \tilde{H} U^{}_{\rm R} + \overline{Q^{}_{\rm L}} Y^{}_{\rm d} H D^{}_{\rm R} + {\rm h.c.} \; ,
\end{eqnarray}
where $Q^{}_{\rm L}$ and $\tilde{H} \equiv i\sigma^{}_2 H^*$ denote the left-handed quark and Higgs doublets, $U^{}_{\rm R}$ and $D^{}_{\rm R}$ are the right-handed quark singlets, $Y^{}_{\rm q}$ for ${\rm q} = {\rm u}$ and ${\rm d}$ stand for the quark mass matrices. After the spontaneous gauge symmetry breaking \cite{Higgs:1964pj}, the quark mass matrices are given by $M^{\prime}_{\rm q} = Y^{}_{\rm q} \langle H \rangle$ with $\langle H \rangle \approx 174~{
\rm GeV}$ being the vacuum expectation value of the Higgs field. Unfortunately, these matrices are not only complex 3 $\times$ 3 structures involving 18 arbitrary parameters each but are also 'weak' basis dependent \cite{Branco:1999nb}. The task of constructing these matrices is therefore critical since the quark mass spectra, flavor mixing angles and CP violation are all determined by these mass matrices. \textit{However, a reasonably modest reconstruction of these matrices from physical observables is expected in a favorable basis, at least within the SM framework, which on one hand translates the observed masses and mixing angles onto the corresponding mass matrices, while on the other hand, avoids the need of additional arbitrary parameters and phases to achieve the same}.  

Due to the absence of flavor-changing right-handed currents in the SM, it is always possible to make the quark mass matrices $M^{\prime}_{\rm u}$ and $M^{\prime}_{\rm d}$ Hermitian, through a unitary transformation of right-handed singlet quark fields in the flavor space.  The quark mass eigenvalues are obtained through the diagonalization of these mass matrices using $V_{\rm u}^\dag M^{\prime}_{\rm u} V_{\rm u} = diag\lbrace {m}_{\rm u}, {m}_{\rm c}, { m}_{\rm t}\rbrace$ and $V_{\rm d}^\dag M^{\prime}_{\rm d} V_{\rm d} = diag\lbrace {m}_{\rm d}, {m}_{\rm s}, { m}_{\rm b}\rbrace$. The resulting quark flavor mixing (CKM) matrix \cite{Cabibbo:1963yz, Kobayashi:1973fv} $V$ arises from the non-trivial mismatch between these diagonalizations i.e. $V=V_{u}^\dag V_d$ and accounts for the flavor changing charged current $W^{\pm}$ interactions through couplings of the form
\begin{equation}\label{eq:chargedint}
- \frac{g}{{\sqrt 2 }}{\overline {\left( {\begin{array}{*{20}{c}}
u&c&t
\end{array}} \right)} _L}{\gamma ^\mu }V{\left( {\begin{array}{*{20}{c}}
d\\
s\\
b
\end{array}} \right)_L}W_\mu ^ +  + h.c.
\end{equation}
Two important observations in this regard need to be emphasized. First, using the running quark masses at the energy scale of $M^{}_Z = 91.2~{\rm GeV}$ \cite{LEUTWYLER1996313, Xing:2011aa, Agashe:2014kda, Olive:2016xmw}: 
\begin{eqnarray}\label{eq:quark masses}
m_u = 1.38^{+0.42}_{-0.41}~{\rm MeV} \; ,~~~~
m_d = 2.82^{+0.48}_{-0.48}~{\rm MeV}\; , \nonumber \\
m_c = 0.638^{+0.043}_{-0.084}~{\rm GeV} \; , ~~
m_s = 57^{+18}_{-12}~{\rm MeV}\;,\nonumber \\
m_t = 172.1^{+1.2}_{-1.2}~{\rm GeV} \; , ~~~~ \; 
m_b = 2.86^{+0.16}_{-0.06}~{\rm GeV}\;, \nonumber \\
m_{u}/m_{d}=0.38-0.58,~~~~~ m_{s}/m_{d}=17-22,
\end{eqnarray}
one observes that the quark mass spectra appear to exhibit a strong hierarchy i.e. ${ m}_{\rm u} \ll {m}_{\rm c} \ll { m}_{\rm t}$ and ${ m}_{\rm d} \ll { m}_{\rm s} \ll { m}_{\rm b}$ and that the hierarchy is relatively stronger in the 'up' quark sector.
Second, the CKM matrix also observes a hierarchical structure viz. $\vert{V_{{\rm{ub}}}}\vert<\vert{V_{{\rm{td}}}}\vert \ll
\vert{V_{{\rm{ts}}}}\vert < \vert{V_{{\rm{cb}}}}\vert\ll \vert{V_{{\rm{cd}}}}\vert < \vert{V_{{\rm{us}}}}\vert<\vert{V_{{\rm{cs}}}}\vert <\vert{V_{{\rm{ud}}}}\vert<\vert{V_{{\rm{tb}}}}\vert$. The current global averages \cite{Agashe:2014kda, Olive:2016xmw} for best fit values of the quark mixing matrix elements are given by 
\begin{widetext}
\begin{equation}\label{PDG best fit}
\left| V \right| = \left( {\begin{array}{*{20}{c}}
{0.97434 \pm 0.00012}&{0.22506 \pm 0.00050}&{0.00357 \pm 0.00015}\\
{0.22492 \pm 0.00050}&{0.97351 \pm 0.00013}&{0.0411 \pm 0.0013}\\
{0.00875\pm 0.00033}&{0.0403\pm 0.0013}&{0.99915 \pm 0.00005}
\end{array}} \right).
\end{equation}
\end{widetext}
The three inner angles of the unitarity triangle are defined through the orthogonality condition: $V^{}_{ud} V^*_{ub} + V^{}_{cd} V^*_{cb} + V^{}_{td} V^*_{tb} = 0$ and are also quite precisely measured \cite{Agashe:2014kda, Olive:2016xmw}, e.g.
\begin{eqnarray}\label{definition inner angles}
\alpha  &\equiv& \arg \left(-\frac{V^{}_{td} V^*_{tb}}{V^{}_{ud} V^*_{ub}}\right)= {87.6^\circ}^{+3.5^\circ}_{-3.3^\circ} \; , \nonumber \\
\beta  &\equiv& \arg \left(-\frac{V^{}_{cd} V^*_{cb}}{V^{}_{td} V^*_{tb}}\right)= {21.8^\circ}^{+0.48^\circ}_{-0.48^\circ} \; , \nonumber \\
\gamma  &\equiv& \arg \left(-\frac{V^{}_{ud} V^*_{ub}}{V^{}_{cd} V^*_{cb}}\right)= {73.2^\circ}^{+6.3^\circ}_{-7.0^\circ} \;
\end{eqnarray}
indicating towards a unitary CKM matrix. Any natural description of the observed flavor mixing should translate these strong hierarchies in the quark mass spectra and the flavor mixing parameters directly onto the hierarchy and phase structure of the corresponding mass matrices \cite{TNYA:TNYA2958, FRITZSCH1977436}. \textit{This paper attempts to address this problem in a clear and simple manner within the SM framework}.

In view of connecting the strong hierarchies in the quark mass spectra and the flavor mixing parameters with corresponding quark mass matrix structures, one successful ansatz incorporating the ‘texture zero’ approach was initiated by Weinberg \cite{TNYA:TNYA2958} and Fritzsch \cite{FRITZSCH1977436} and this approach has been widely adopted for understanding the behavior of flavor mixing patterns both in the quark \cite{Branco:1999nb, FRITZSCH20001, PhysRevD.76.033008, Gupta:2009ur, Verma:2009gf, Mahajan:2009wd, Verma:2013qta, Sharma:2014tea, Ludl:2014axa, Sharma:2015gfa, Verma:2015mgd, Fritzsch:2015foa, Giraldo:2015cpp, Xing:2015sva, Singh:2016qcf, Ahuja:2016khy, Tanimoto:2016rqy, Fritzsch:2017tyf, Ahuja:2017fhi} as well as the lepton sectors \cite{BRANCO2009340, Ahuja:2009jj, Verma:2010jy, Verma:2013cza, Fakay:2013gf, Verma:2014woa, Verma:2014lpa, Ludl:2015lta, Chen:2016ica, Verma:2016qhy, Dziewit:2016qri, Kumar:2017hjn, Han:2017wnk}. A particular texture structure is said to be texture 'n' zero, if it has 'n' number of non-trivial zeros, for example, if the sum of the number of diagonal zeros and half the number of the symmetrically placed
off diagonal zeros is 'n'. Essentially, these texture zeros are phenomenological zeros that represent entries in the mass matrix which are highly suppressed as compared to their nearest neighbors. In the texture-zero approach to quark mass matrices, recently, it has been widely discussed \cite{Sharma:2014tea, Sharma:2015gfa, Ahuja:2016khy, Ahuja:2017fhi} that only a certain class of Hermitian texture zero structures, namely the Fritzsch-like texture 4-zeros, are viable in the quark sector with q = u,d. 
\begin{equation}
M^{\prime}_{q}=\left( {\begin{array}{*{20}{c}}
\bf{0}&{{a_{\rm{q}}}}&\bf{0}\\
{{a^{*}_{\rm{q}}}}&{{d_{\rm{q}}}}&{{b_{\rm{q}}}}\\
\bf{0}&{{b^{*}_q}}&{{c_{\rm{q}}}}
\end{array}} \right)
\end{equation}

These however require at least 10 (equal to number of physical observables) matrix parameters including two non-trivial phases \cite{Fritzsch:2002ga} to account for the observed flavor mixings. In addition, the relations for the mixing parameters in such scenarios are quite complicated \cite{Verma:2009gf, Verma:2013qta} and difficult to comprehend. In the light of these recent works \cite{Sharma:2014tea, Sharma:2015gfa, Ahuja:2016khy, Ahuja:2017fhi}, it becomes desirable to investigate if a more generic approach using fundamental characteristics of CKM matrix and measured hierarchies of physical observables allow for a modest reconstruction of the texture based quark mass matrices accounting for the observed flavor mixing.

The structure of the paper is as follows. Section-II uses the Wolfenstein parametrization for $V$ to gain some insight on the gross structural and hierarchical features of these mass matrices from a general point of view. Later, in Section-III, the standard parametrization for $V$ is used along with the hierarchical nature of quark masses and mixing angles to extract vital clues on the flavor mixing parameters along with the CP phase and the three angles of the unitarity triangle. Section-IV attempts to translate this information to reconstruct a modest structure for the corresponding mass matrices through a leading order as well an exact analytic solution. This is followed by conclusions in Section-V.
\section{Wolfenstein Parametrization}
The Wolfenstein parametrization \cite{Wolfenstein:1983yz} for $V$ is instructive in providing vital clues towards the formulation of these quark mass matrices, e.g.
\begin{widetext}
\begin{equation}\label{Wolfenstein}
V = \left( {\begin{array}{*{20}{c}}
{1 - {{{\lambda ^2}} \mathord{\left/
 {\vphantom {{{\lambda ^2}} 2}} \right.
 \kern-\nulldelimiterspace} 2}}&\lambda &{A{\lambda ^3}\left( {\rho  - i\eta } \right)}\\
{ - \lambda }&{1 - {{{\lambda ^2}} \mathord{\left/
 {\vphantom {{{\lambda ^2}} 2}} \right.
 \kern-\nulldelimiterspace} 2}}&{A{\lambda ^2}}\\
{A{\lambda ^3}\left( {1 - \rho  - i\eta } \right)}&{ - A{\lambda ^2}}&1
\end{array}} \right)\simeq \left( {\begin{array}{*{20}{c}}
1&\lambda &{0.3{\lambda ^3}}\\
{ - \lambda }&1&{ 0.8{\lambda ^2}}\\
{0.6{\lambda ^3}}&{ - 0.8{\lambda ^2}}&1
\end{array}} \right),
\end{equation} 
\end{widetext}
where $\lambda  = 0.22$, ${\rm{A = 0}}{\rm{.82}}$, $\rho  = {\rm{0}}{\rm{.13}}$ and $\eta {\rm{ = 0}}{\rm{.345}}$ have been used \cite{Agashe:2014kda} and the phases have been ignored for simplicity. In the $u$-diagonal flavor basis
\begin{eqnarray}
{M^{\prime}_{\rm{d}}}=V M_{\rm{d}}^{{\rm{Diag}}}{V^\dag }\;  \nonumber\\
\sim \left( {\begin{array}{*{20}{c}}
{{m_{\rm{d}}}}&{\sqrt {{m_{\rm{d}}}{m_{\rm{s}}}} }&{0.1\sqrt {{m_{\rm{d}}}{m_{\rm{b}}}} }\\
{ - \sqrt {{m_{\rm{d}}}{m_{\rm{s}}}} }&{{m_{\rm{s}}}}&{0.3\sqrt {{m_{\rm{s}}}{m_{\rm{b}}}} }\\
{0.1\sqrt {{m_{\rm{d}}}{m_{\rm{b}}}} }&{ - 0.3\sqrt {{m_{\rm{s}}}{m_{\rm{b}}}} }&{{m_{\rm{b}}}}
\end{array}} \right).\;  
\end{eqnarray}
Using $M_{\rm{d}}^{{\rm{Diag}}}=diag\lbrace -0.0031, 0.057, 2.910\rbrace GeV$, the following numerical estimate can be made, e.g.
\begin{equation}\label{real md}
|M^{\prime}_{\rm d}| \simeq \left( {\begin{array}{*{20}{c}}
0.00001&0.01&0.01\\
0.01&0.058&0.12\\
0.01&0.12&2.90
\end{array}} \right){\rm{ GeV}}.
\end{equation}
Interestingly, the above matrix is hierarchical among the diagonal elements with ${\bf{0}}\simeq(M^{\prime}_{\rm d})_{11}\ll m_{\rm d}\ll (M^{\prime}_{\rm d})_{12,21}\simeq \sqrt{m_{\rm d}m_{\rm s}}\sim(M^{\prime}_{\rm d})_{13,31}<(M^{\prime}_{\rm d})_{22}\simeq m_{\rm s}\ll(M^{\prime}_{\rm d})_{33}\simeq m_{\rm b}$. However, we should also require some insight on the possible phase structure of $M^{\prime}_{\rm d}$.
\section{Standard Parametrization}
Whereas, the Wolfenstein parametrization can be used to understand the hierarchy of the mass matrix elements, the standard parametrization \cite{Agashe:2014kda} for $V$ allows to extract several vital clues for the flavor mixing parameters as well as the three angles of the unitarity triangle. Most importantly, it also facilitates the mass matrices in attaining a minimal phase structure as discussed below. In this parametrization, $V$ is expressed in terms of the three mixing angles $s_{ij}=\rm{sin}~ \theta_{ij}$, $c_{ij}=\rm{cos}~ \theta_{ij}$ where $\theta_{ij}$ $(i,j=1,2,3)$ lie in the first quadrant i.e. $s_{ij},c_{ij}\geq 0$ and a CP-violation phase $\delta_{13}$ associated with the flavor-changing processes in the SM appears, e.g.
\begin{widetext}
\begin{eqnarray}\label{choice of parametrization}
V = \left( {\begin{array}{*{20}{c}}
1&0&0\\
0&{{c_{23}}}&{{s_{23}}}\\
0&{ - {s_{23}}}&{{c_{23}}}
\end{array}} \right)\left( {\begin{array}{*{20}{c}}
{{c_{13}}}&0&{{s_{13}}{e^{  -i{\delta _{13}}}}}\\
0&{ 1}&0\\
{ - {s_{13}}{e^{i{\delta _{13}}}}}&0&{{c_{13}}}
\end{array}} \right)\left( {\begin{array}{*{20}{c}}
{{c_{12}}}&{{s_{12}}}&0\\
{ - {s_{12}}}&{{c_{12}}}&0\\
0&0&{  1}
\end{array}} \right)\; \nonumber\\
= \left( {\begin{array}{*{20}{c}}
{{c_{12}}{c_{13}}}&{{s_{12}}{c_{13}}}&{  {s_{13}}{e^{  -i{\delta _{13}}}}}\\
{-{s_{12}}{c_{23}} - {c_{12}}{s_{23}}{s_{13}}{e^{i{\delta _{13}}}}}&{ {c_{12}}{c_{23}} - {s_{12}}{s_{23}}{s_{13}}{e^{i{\delta _{13}}}}}&{  {s_{23}}{c_{13}}}\\
{ {s_{12}}{s_{23}}-{s_{13}}{c_{12}}{c_{23}}{e^{i{\delta _{13}}}}  }&{-{s_{23}}{c_{12}} - {s_{12}}{s_{13}}{c_{23}}{e^{i{\delta _{13}}}}}&{  {c_{23}}{c_{13}}}
\end{array}} \right)
= \left( {\begin{array}{*{20}{c}}
V_{ud}& V_{us}& V_{ub}\\
V_{cd}& V_{cs}& V_{cb}\\
V_{td}& V_{ts}& V_{tb}
\end{array}} \right).
\end{eqnarray}
\end{widetext}
Using $s_{13}\ll s_{23}\ll s_{12}$, the above structure can be reduced to the following form without losing generality,
\begin{equation}\label{reduced CKM}
V=\left( {\begin{array}{*{20}{c}}
{{c_{12}}}&{{s_{12}}}&{  {s_{13}}{e^{  -i{\delta _{13}}}}}\\
{-{s_{12}}{c_{23}}}&{{c_{12}}{c_{23}}}&{ {s_{23}}}\\
{ {s_{12}}{s_{23}}- {s_{13}}{c_{12}}{e^{i{\delta _{13}}}} }&{-{s_{23}}{c_{12}}}&{  {c_{23}}}
\end{array}} \right)
\end{equation}
and follows from $s_{12}s_{23}s_{13}\ll 1$, $s_{12}s_{13}\ll s_{23}$, $s_{23}s_{13}\ll s_{12}$, $s_{12}s_{23}\sim s_{13}$ and $c_{ij}\simeq 1$. The Eq.~(\ref{reduced CKM}) also provides a natural explanation for $|V_{\rm {cd}}|<|V_{\rm {us}}|$, $|V_{\rm {ts}}|<|V_{\rm {cb}}|$ and $|V_{ud}||V_{tb}|=|V_{cs}|$ which can be verified using the best-fit values in Eqn.(\ref{PDG best fit}). This further allows to establish trivial relations for the unitarity angles 
\begin{equation}\label{beta relation}
\beta  =  - \arg \left( {1 - \frac{{{s_{13}}{c_{12}}}}{{{s_{12}}{s_{23}}}}{e^{ i{\delta_{13}}}}} \right).
\end{equation}
The above relation provides quite an accurate prediction for $\beta$ calculation using Eqns.(\ref{PDG best fit}) and (\ref{definition inner angles}) with $\delta_{13}\simeq \gamma$. Likewise, the Jarlskog's CP invariant parameter \cite{PhysRevLett.55.1039, Jarlskog:1985cw, Wu:1985ea} $J_{\rm{CP}}=Im[V_{us}V_{cb}V^{*}_{cs}V^{*}_{ub}]\simeq {c_{12}}s_{12}s_{23}s_{13}{\rm {sin}}\delta_{13}$. The current global average \cite{Agashe:2014kda, Olive:2016xmw} for this parameter is given by ${J_{{\rm{CP}}}} = (3.04_{ - 0.20}^{ + 0.21}) \times {10^{ - 5}}$. Furthermore, since there is only one physical phase $\delta_{13}$ in the CKM matrix, it is desirable to translate this single phase onto the corresponding mass matrices.  To this end, one may always start from a basis wherein one of the matrices $M_{\rm{q}}$ ($\rm{q}=u,d$) is real diagonal and the other is an arbitrary Hermitian mass matrix \cite{Branco:1999nb, BRANCO2009340}. Noting that the hierarchy is much stronger in the 'up'  quark masses, one is compelled to consider the diagonal flavor basis of $M^{\prime}_{\rm u}$, such that
\begin{equation}\label{u-diagonal Md}
M^{\prime}_{\rm d}=VM_{\rm{d}}^{{\rm{Diag}}}{V^{\dag}}
\end{equation}
describes the matrix elements of $M^{\prime}_{\rm d}$ in terms of the physical observables namely the three quark masses and four mixing parameters. However, the structure for the CKM matrix V in Eqn.(\ref{choice of parametrization}), when substituted in Eqn.(\ref{u-diagonal Md}), leads to mathematically complicated relations between the matrix elements of $M^{\prime}_{\rm d}$ and the physical observables in V which are difficult to comprehend and a deeper insight into such relationship cannot be established. In particular, it is unclear if the three overall phases associated with the off-diagonal mass matrix elements establish any trivial relationship to $\delta_{13}$ in CKM matrix. Explicitly, with $M^{Diag}_{\rm{d}}=(-m_{\rm{d}},m_{\rm{s}},m_{\rm{b}})$ and substituting a relatively simpler Eq.(\ref{reduced CKM}) in relation (\ref{u-diagonal Md}), one obtains
\begin{equation}
\begin{array}{c}
M^{\prime}_{\rm{d}}(12)=c_{12}c_{23}(m_{\rm{d}}+m_{\rm{s}})s_{12} + m_{\rm{b}}s_{13}s_{23}e^{-i\delta_{13}} = a_{\rm{d}}e^{-i\alpha_{\rm{d}}},\\
M^{\prime}_{\rm{d}}(13)=c_{23}s_{13}m_{\rm{b}}e^{-i\delta_{13}}+c^{2}_{12}s_{13}m_{\rm{d}}e^{-i\delta_{13}}-c_{12}s_{12}s_{23}(m_{\rm{d}}+m_{\rm{s}})=f_{\rm{d}}e^{-i\delta_{\rm{d}}},\\
M^{\prime}_{\rm{d}}(23)=s_{23}c_{23}m_{\rm{b}}-c^{2}_{12}s_{23}c_{23}m_{\rm{s}}+m_{\rm{d}}s_{12}c_{23}(s_{12}s_{23}-s_{13}c_{12}e^{-i\delta_{13}}) = b_{\rm{d}}e^{-i\beta_{\rm{d}}}.
\end{array}
\end{equation} 

In the following section, we show that the observed strong hierarchies in the quark mass spectra and mixing angles can be of immense help in deducing modest relations among these parameters as well as implications for $\alpha_{\rm{d}}, \beta_{\rm{d}}$ and $\delta_{\rm{d}}$ viz-a-viz $\delta_{13}$ leading to a rather simplistic structure for the mass matrices. 
\section{Diagonal Flavor Basis}
\subsection{Leading Order Solution}
In the flavor basis of diagonal $M^{\prime}_{\rm u}=diag\lbrace m_{u},m_{c},m_{t}\rbrace$ and using observed hierarchies in quark masses and mixing angles along with $M_{\rm{d}}^{{\rm{Diag}}}=diag\lbrace -m_{\rm d}, m_{\rm s}, { m}_{\rm b}\rbrace$, one obtains the following modest structure for $M^{\prime}_{\rm d}$ to a leading order, e.g.
\begin{widetext}
\begin{equation}\label{clues to exact md}
{M^{\prime}_{\rm{d}}} \simeq \left( {\begin{array}{*{20}{c}}
{m_{\rm{s}}}{s^2_{12}}-{m_{\rm{d}}} &{({m_{\rm{s}}}+{m_{\rm{d}}}){s_{12}}}&{{m_{\rm{b}}}{s_{13}}{e^{-i{\delta _{13}}}}}\\
{({m_{\rm{s}}}+{m_{\rm{d}}}){s_{12}}}&{{m_{\rm{b}}}s_{23}^2 + {m_{\rm{s}}}}&{({m_{\rm{b}}}-{m_{\rm{s}}}){s_{23}}}\\
{{m_{\rm{b}}}{s_{13}}{e^{i{\delta _{13}}}}}&{({m_{\rm{b}}}-{m_{\rm{s}}}){s_{23}}}&{m_{\rm{b}}}c_{23}^2
\end{array}} \right)= \left( {\begin{array}{*{20}{c}}
\bf{0}&{{a_{\rm{d}}}}{e^{ - i{\alpha _{\rm{d}}}}}&{{f_{\rm{d}}}{e^{ - i{\delta _{\rm{d}}}}}}\\
{{a_{\rm{d}}}}{e^{ i{\alpha _{\rm{d}}}}}&{{d_{\rm{d}}}}&{{b_{\rm{d}}}}{e^{ - i{\beta _{\rm{d}}}}}\\
{{f_{\rm{d}}}{e^{ i{\delta _{\rm{d}}}}}}&{{b_d}}{e^{ i{\beta _{\rm{d}}}}}&{{c_{\rm{d}}}}
\end{array}} \right).
\end{equation}
\end{widetext}
where $\alpha_{\rm{d}}=\beta_{\rm{d}}=0$, and hence $a_{\rm{d}}$, $b_{\rm{d}}$, $f_{\rm{d}}$, $d_{\rm{d}}$ and $c_{\rm{d}}$ can be inferred to be purely real within an error of less than a percent due to the strong hierarchy in the quark sector.  The modest relationship among the matrix elements of $M^{\prime}_{\rm d}$ and the physical observables: $m_{\rm d}$, $m_{\rm s}$, $m_{\rm b}$, $s_{12}$, $s_{13}$, $s_{23}$ and $\delta_{13}$ is clearly manifest, i.e.
\begin{equation}\label{modest relations}
\begin{array}{c}
a_{\rm{d}}=s_{12}(m_{\rm{d}}+m_{\rm{s}}),\\
b_{\rm{d}}=s_{23}(m_{\rm{b}}-m_{\rm{s}}),\\
f_{\rm{d}}=s_{13} m_{\rm{b}},\\
\delta_{\rm{d}}=\delta_{13}.
\end{array}
\end{equation}
Note that $M^{\prime}_{\rm d}(11) ={m_{\rm{s}}}{s^2_{12}}-{m_{\rm{d}}}\leq 4\times 10^{-4}=\bf{0}$ in Eqs.~(\ref{real md}) and (\ref{clues to exact md}) is a natural consequence of the precision measured values of $m_{\rm{d}}$, $m_{\rm{d}}$ and $s_{12}=\mid V_{us}\mid$ and are also consistent with the Gatto-Sartori-Tonin (GST) relation \cite{Gatto:1968ss} $s_{12}\simeq \sqrt{m_{\rm d}/m_{\rm s}}$. This is emphasized in Fig.1, depicted through histogram of $M^{\prime}_{\rm d}(1,1)$ obtained for all running values of the quark masses and observed $V_{us}$ values consistent with Eqs.(\ref{eq:quark masses}) and (\ref{PDG best fit}). 
\begin{figure}
\includegraphics[scale=1.0]{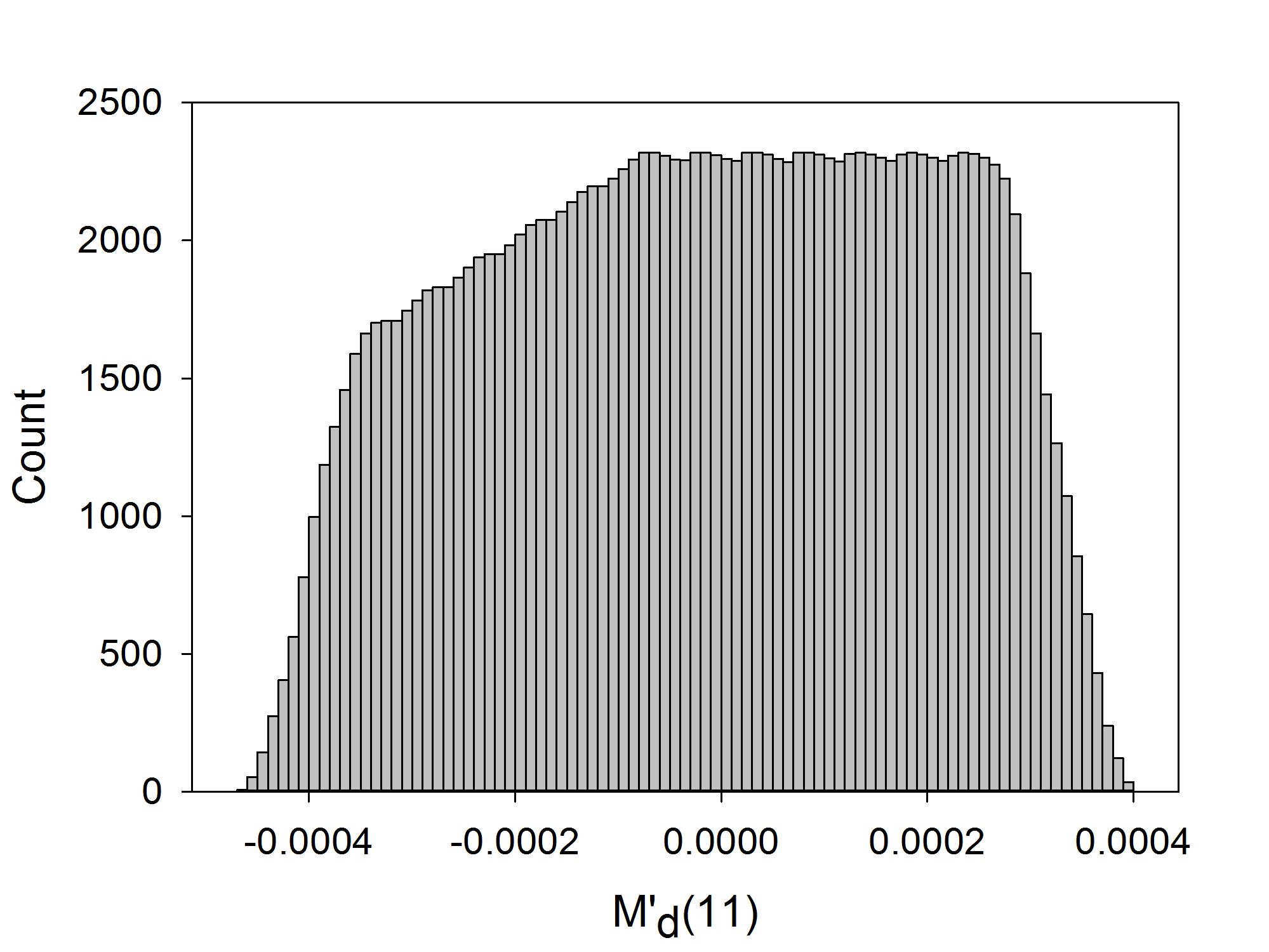}
\caption{Histogram for allowed values of $M^{\prime}_{\rm d}(11)$.}
\end{figure}

The absence of complex phases with other off-diagonal matrix elements of $M^{\prime}_{\rm d}$ in Eq.~(\ref{clues to exact md}) follows naturally from the hierarchical and phase structure of $V$ in Eq.~(\ref{reduced CKM}). Clearly the mass matrix in Eqn.(\ref{clues to exact md}) associates the CP-violating phase in the CKM matrix predominantly with the off-diagonal $M^{\prime}_{\rm d}(1,3)$ and $M^{\prime}_{\rm d}(3,1)$ elements. \textit{This makes it vital to investigate whether an exact analytic solution to the the above leading order texture 4-zero matrices is also in agreement with the current quark flavor mixing and CP-violation data within the known precision limits.}
\subsection{Exact Solution}
A possible analytic solution of the above $M^{\prime}_{\rm d}$ texture structure can be achieved through a complex unitary transformation $R$ on real symmetric $M_{\rm d}$ through 
\begin{widetext}
\begin{equation}\label{md reconstruction}
\begin{array}{r}
{M^{\prime}_{\rm{d}}} = {R^\dag }{M_{\rm{d}}}R= \left( {\begin{array}{*{20}{c}}
1&0&{{s_{\rm{d}}}{e^{-i{\delta _{\rm{d}}}}}}\\
0&1&0\\
{ - {s_{\rm{d}}}{e^{i{\delta _{\rm{d}}}}}}&0&1
\end{array}} \right)\left( {\begin{array}{*{20}{c}}
{{e^{\prime}_{\rm{d}}}}&{{a_{\rm{d}}}}&0\\
{{a_{\rm{d}}}}&{{d_{\rm{d}}}}&{{b_{\rm{d}}}}\\
0&{{b_{\rm{d}}}}&{{c_{\rm{d}}}}
\end{array}} \right)\left( {\begin{array}{*{20}{c}}
1&0&{ - {s_{\rm{d}}}{e^{-i{\delta _{\rm{d}}}}}}\\
0&1&0\\
{{s_{\rm{d}}}{e^{ i{\delta _{\rm{d}}}}}}&0&1
\end{array}} \right)\\ 
\simeq \left( {\begin{array}{*{20}{c}}
\bf{0}&{{a_{\rm{d}}}}&{{f_{\rm{d}}}{e^{ - i{\delta _{\rm{d}}}}}}\\
{{a_{\rm{d}}}}&{{d_{\rm{d}}}}&{{b_{\rm{d}}}}\\
{{f_{\rm{d}}}{e^{ i{\delta _{\rm{d}}}}}}&{{b_d}}&{{c_{\rm{d}}}}
\end{array}} \right)
\end{array}
\end{equation}
\end{widetext}
where $s_{\rm d}=sin\theta_{\rm d}=\sqrt{-e^{\prime}_{\rm d}/c_{\rm d}}\ll 1$, $cos\theta_{\rm d}\simeq 1$ and $-m_{\rm d}<e^{\prime}_{\rm d}<0$ follows from $M^{\prime}_{\rm d}(1,1)=\bf{0}$ for conformity with the leading order solution. The CKM matrix then originates from $V=R^{\dag}O_{\rm d}$ where $O_{\rm d}$ is the orthogonal transformation that diagonalizes $M_{\rm d}$ through $O^{T}_{\rm d}M_{\rm d}O_{\rm d}= diag\lbrace {km}_{\rm d}, -k{m}_{\rm s}, {m}_{\rm b}\rbrace$. Note that $M^{\prime \rm{Diag}}_{\rm{d}}=M_{\rm{d}}^{{\rm{Diag}}}=diag\lbrace km_{\rm d}, -km_{\rm s}, { m}_{\rm b}\rbrace$ follows from the unitary $R$ transformation and $k=\lbrace-1,+1\rbrace$ accommodates a negative mass eigenvalue associated with $m_{\rm d}$ or $m_{\rm s}$, respectively.  The above mass matrix involves only three non-trivial parameters namely $e^{\prime}_{\rm d}$, $d_{\rm d}$ and $\delta_{\rm d}$ (in addition to the three quark masses). In particular, $O_{\rm d}$ is expressed as \cite{Verma:2013qta}
\begin{widetext}
\begin{equation}
{O_{\rm{d}}} = \left( {\begin{array}{*{20}{c}}
{\sqrt {\frac{{(k{e^{\prime}_{\rm{d}}} + {m_{\rm s}})({m_{\rm b}} - {e^{\prime}_{\rm{d}}})({c_{\rm{d}}} - k{m_{\rm d}})}}{{({c_{\rm{d}}} - {e^{\prime}_{\rm{d}}})({m_{\rm b}} - k{m_{\rm d}})({m_{\rm s}} + {m_{\rm d}})}}} }&{\sqrt {\frac{{({m_{\rm d}} - k{e^{\prime}_{\rm{d}}})({m_{\rm b}} - {e^{\prime}_{\rm{d}}})({c_{\rm{d}}} + k{m_{\rm s}})}}{{({c_{\rm{d}}} - {e^{\prime}_{\rm{d}}})({m_{\rm b}} + k{m_{\rm s}})({m_{\rm s}} + {m_{\rm d}})}}} }&{\sqrt {\frac{{({m_{\rm d}} - k{e^{\prime}_{\rm{d}}})(k{e^{\prime}_{\rm{d}}} + {m_{\rm s}})({m_{\rm b}} - {c_{\rm{d}}})}}{{({c_{\rm{d}}} - {e^{\prime}_{\rm{d}}})({m_{\rm b}} + k{m_{\rm s}})({m_{\rm b}} - k{m_{\rm d}})}}} }\\
{k\sqrt {\frac{{({m_{\rm d}} - k{e^{\prime}_{\rm{d}}})({c_{\rm{d}}} - k{m_{\rm d}})}}{{({m_{\rm b}} - k{m_{\rm d}})({m_{\rm s}} + {m_{\rm d}})}}} }&{ - k\sqrt {\frac{{(k{e^{\prime}_{\rm{d}}} + {m_{\rm s}})({c_{\rm{d}}} + k{m_{\rm s}})}}{{({m_{\rm b}} + k{m_{\rm s}})({m_{\rm s}} + {m_{\rm d}})}}} }&{\sqrt {\frac{{({m_{\rm b}} - {e^{\prime}_{\rm{d}}})({m_{\rm b}} - {c_{\rm{d}}})}}{{({m_{\rm b}} + k{m_{\rm s}})({m_{\rm b}} - k{m_{\rm d}})}}} }\\
{ - k\sqrt {\frac{{({m_{\rm d}} - k{e^{\prime}_{\rm{d}}})({m_{\rm b}} - {c_{\rm{d}}})({c_{\rm{d}}} + k{m_{\rm s}})}}{{({c_{\rm{d}}} - {e^{\prime}_{\rm{d}}})({m_{\rm b}} - k{m_{\rm d}})({m_{\rm s}} + {m_{\rm d}})}}} }&{k\sqrt {\frac{{(k{e^{\prime}_{\rm{d}}} + {m_{\rm s}})({c_{\rm{d}}} - k{m_{\rm d}})({m_{\rm b}} - {c_{\rm{d}}})}}{{({c_{\rm{d}}} - {e^{\prime}_{\rm{d}}})({m_{\rm b}} + k{m_{\rm s}})({m_{\rm s}} +{m_{\rm d}})}}} }&{\sqrt {\frac{{({m_{\rm b}} - {e^{\prime}_{\rm{d}}})({c_{\rm{d}}} - k{m_{\rm d}})({c_{\rm{d}}} + k{m_{\rm s}})}}{{({c_{\rm{d}}} - {e^{\prime}_{\rm{d}}})({m_{\rm b}} + k{m_{\rm s}})({m_{\rm b}} - k{m_{\rm d}})}}} }
\end{array}} \right),
\end{equation}
\end{widetext}
\begin{eqnarray}
{c_{\rm{d}}} = k{m_{\rm d}} - k{m_{\rm s}} + {m_{\rm b}} - {d_{\rm{d}}} - {e^{\prime}_{\rm{d}}}\; , \nonumber\\
{a_{\rm{d}}} = \sqrt {\frac{{\left( {{m_{\rm d}} - k{e^{\prime}_{\rm{d}}}} \right)\left( {{m_{\rm s}} + k{e^{\prime}_{\rm{d}}}} \right)\left( {{m_{\rm b}} - {e^{\prime}_{\rm{d}}}} \right)}}{{\left( {{c_{\rm{d}}} - {e^{\prime}_{\rm{d}}}} \right)}}} \; , \nonumber \\
{b_{\rm{d}}} = \sqrt {\frac{{\left( {{c_{\rm{d}}} - k{m_{\rm d}}} \right)\left( {{c_{\rm{d}}} + k{m_{\rm s}}} \right)\left( {{m_{\rm b}} - {c_{\rm{d}}}} \right)}}{{\left( {{c_{\rm{d}}} - {e^{\prime}_{\rm{d}}}} \right)}}} \; , \nonumber\\
({m_{\rm b}} - {m_{\rm s}} - {e^{\prime}_{\rm{d}}}) > {d_{\rm{d}}} > ({m_{\rm d}} - {m_{\rm s}} - {e^{\prime}_{\rm{d}}})\;, \nonumber\\
0 > {e^{\prime}_{\rm{d}}} >  - {m_{\rm d}}\;.
\end{eqnarray}
Using ${{{m}}_{\rm{d}}}\ll{{{m}}_{\rm{s}}}\ll{{{m}}_{\rm{b}}}$, a further simplification can be achieved by a redefinition of the parameters $e^{\prime}_{\rm d}$ and $d_{\rm d}$ as  
\begin{eqnarray}\label{hierarchy parameters hp}
{{{{\xi }}_{\rm{d}}}}=e^{\prime}_{\rm d}/m_{\rm d}\;, ~~
{{{{\zeta }}_{\rm{d}}}}=d_{\rm d}/c_{\rm d},\;
\end{eqnarray}
such that $|{\xi _{\rm{d}}}|\ll 1,~~|{\zeta _{\rm{d}}}|\ll 1$. The resulting mixing matrix $V=R^{\dag}O_{\rm d}$, with an error of less than a percent, is expressed below
\begin{widetext}
\begin{equation}\label{exact ckm}
V \simeq \left( {\begin{array}{*{20}{c}}
{{\sigma _{\rm{d}}}}&{{\sigma _{\rm{d}}}\sqrt {\frac{{{m_{\rm{d}}}}}{{{m_{\rm{s}}}}}} }&{\frac{{{m_{\rm{s}}}}}{{{m_{\rm{b}}}}}{s_{12}}{s_{23}}+\sqrt {\frac{{{m_{\rm{d}}}}}{{{m_{\rm{b}}}}}} \sqrt { - {\xi _{\rm{d}}}} {e^{ - i{\delta _{\rm{d}}}}}}\\
{ k{\sigma _{\rm{d}}}\sqrt {\frac{{{m_{\rm{d}}}}}{{{m_{\rm{s}}}}}} \sqrt {1 - {\Delta _{\rm{d}}}} }&{{-k\sigma _{\rm{d}}}\sqrt {1 - {\Delta _{\rm{d}}}} }&\sqrt {\frac{{{\zeta _{\rm{d}}} + k{\Delta _{\rm{d}}}}}{{1 + {\zeta _{\rm{d}}}}}} \\
{\left( -k\sigma_{\rm d}{\sqrt {\frac{{{m_{\rm{d}}}}}{{{m_{\rm{s}}}}}} \sqrt {{\zeta _{\rm{d}}} + {k\Delta _{\rm{d}}}}  - {e^{i{\delta _{\rm{d}}}}}{\sigma _{\rm{d}}}\sqrt {\frac{{{m_{\rm{d}}}}}{{{m_{\rm{b}}}}}} \sqrt { - {\xi _{\rm{d}}}} } \right)}&k{\sigma _{\rm{d}}}\sqrt {\frac{{{\zeta _{\rm{d}}} + k{\Delta _{\rm{d}}}}}{{1 + {\zeta _{\rm{d}}}}}} &{\sqrt {1 - {\Delta _{\rm{d}}}} }
\end{array}} \right),
\end{equation}
\end{widetext}
where
\begin{equation}\label{definition 1}
{\sigma _{\rm{d}}} \simeq \sqrt{\frac{m_{\rm s}}{m_{\rm d} + m_{\rm s}}},~~~~
{{\Delta _{\rm{d}}} = \frac{{{m_{\rm{s}}} - {m_{\rm{d}}}}}{{{m_{\rm{b}}}}}}
\end{equation}
have been introduced for notation convenience. Note that, for $k=-1$ there is a complete agreement of the above structure of $V$ with that of Eq.~(\ref{reduced CKM}) in the context of the (-ve) signs associated with $V_{cd}$, $V_{cs}$, $V_{td}$ and $V_{ts}$ and a one to one correspondence is then clearly evident through the relations
\begin{eqnarray}\label{predictions}
|V_{ud}|=c_{12}= \sigma_{\rm d}= \sqrt{\frac{m_{\rm s}}{m_{\rm d} + m_{\rm s}}}\; , \nonumber\\
|V_{us}|=s_{12}=\sqrt{m_{\rm d}/(m_{\rm d}+m_{\rm s})}\; , \nonumber\\
|V_{ub}|={s_{13}}\simeq {s_{\rm{d}}} =\sqrt { - {\xi _{\rm{d}}}} \sqrt {\frac{{{m_{\rm{d}}}}}{{{m_{\rm{b}}}}}}\; , \nonumber\\
|V_{cd}|=|V_{us}||V_{tb}|<|V_{us}|\; , \nonumber\\
|V_{cs}|=|V_{ud}||V_{tb}|\; , \nonumber\\
|V_{cb}|=s_{23}=\sqrt{\zeta_{\rm d}-\Delta_{\rm d}}\; , \nonumber\\
V_{td}=s_{12}s_{23}-c_{12}s_{13}e^{i\delta_{\rm{d}}}\; , \nonumber\\
|V_{ts}|=|V_{ud}||V_{cb}|<|V_{cb}|\; , \nonumber\\
|V_{tb}|=c_{23}=\sqrt{(m_{\rm b}-m_{\rm s}+m_{\rm d})/m_{\rm b}}\;, \nonumber\\\;  \nonumber\\
{\delta _{13}} = {\tan ^{ - 1}}\left( {\frac{{{m_{\rm{b}}}{s_{13}}\sin {\delta _{\rm{d}}}}}{{{m_{\rm{s}}}{s_{12}}{s_{23}} + {m_{\rm{b}}}{s_{13}}\cos {\delta _{\rm{d}}}}}} \right)\simeq \delta_d\; , \nonumber\\
\beta  = - \arg \left( {1 - {\sigma _{\rm{d}}}\sqrt {\frac{{ - {m_{\rm{s}}}{\xi _{\rm{d}}}}}{{{m_{\rm{b}}}\left( {{\zeta _{\rm{d}}} - {\Delta _{\rm{d}}}} \right)}}} {e^{ i{\delta _{\rm{d}}}}}} \right).
\end{eqnarray}
The corresponding  $M^{\prime}_{\rm d}$ in terms of $\xi_{\rm d}$, $\zeta_{\rm d}$, $\delta_{\rm d}$ and the quark masses appears below
\begin{widetext}
\begin{eqnarray}\label{exact md}
M_{\rm{d}}^\prime  \simeq \left( {\begin{array}{*{20}{c}}
\bf{0}&{\sqrt {{m_{\rm{d}}}{m_{\rm{s}}}} }&\sqrt { - {\xi _{\rm{d}}}{m_{\rm{d}}}{m_{\rm{b}}}}{e^{-i{\delta _{\rm{d}}}}}\\
{\sqrt {{m_{\rm{d}}}{m_{\rm{s}}}} }&{m_{\rm{b}}}{\zeta _{\rm{d}}}&{{{m_{\rm{b}}}\sqrt {{\zeta _{\rm{d}}} - {\Delta _{\rm{d}}}} } \mathord{\left/
 {\vphantom {{{m_{\rm{b}}}\sqrt {{\zeta _{\rm{d}}} - {\Delta _{\rm{d}}}} } {\left( {1 + {\zeta _{\rm{d}}}} \right)}}} \right.
 \kern-\nulldelimiterspace} {\left( {1 + {\zeta _{\rm{d}}}} \right)}}\\
\sqrt { - {\xi _{\rm{d}}}{m_{\rm{d}}}{m_{\rm{b}}}}{e^{i{\delta _{\rm{d}}}}}&{{{m_{\rm{b}}}\sqrt {{\zeta _{\rm{d}}} - {\Delta _{\rm{d}}}} } \mathord{\left/
 {\vphantom {{{m_{\rm{b}}}\sqrt {{\zeta _{\rm{d}}} - {\Delta _{\rm{d}}}} } {\left( {1 + {\zeta _{\rm{d}}}} \right)}}} \right.
 \kern-\nulldelimiterspace} {\left( {1 + {\zeta _{\rm{d}}}} \right)}}&m_{\rm{b}}(1-\zeta_{\rm{d}}+\Delta_{\rm{d}})
\end{array}} \right),\;\nonumber \\
\end{eqnarray}
\end{widetext}
where 
\begin{equation}
\vert\xi_{\rm d}\vert=m_{b}s^2_{13}/m_d,~~
\zeta_{\rm d}=s^2_{23}+\Delta_{\rm{d}}, ~~
\delta_{\rm{d}}\simeq \delta_{13}.
\end{equation}
The above matrix is in remarkable agreement with the leading order result in Eq.(\ref{clues to exact md}), establishing that a particular class of texture 4-zero quark mass matrix in Eq.(\ref{clues to exact md}) follows directly from the observed precision measurements on physical observables in a suitable basis.

We now use the available mixing data for the down sector quark masses along with four flavor mixing parameters $|V_{us}|$,$|V_{cb}|$,$|V_{ub}|$ and $\beta$ at 1$\sigma$ level from Eqs.(\ref{eq:quark masses})-(\ref{definition inner angles}) for the analysis. The best-fit values for the six free parameters $\xi_{\rm d}$, $\zeta_{\rm d}$, $\delta_{\rm d}$, $m_{\rm d}$, $m_{\rm s}$ and $m_{\rm b}$ in the exact solution case are presented below: 
\begin{eqnarray}
{m_{\rm{d}}} = 3.08{\rm{MeV}},~~ {m_{\rm{s}}} = 57.0{\rm{MeV}},~~{m_{\rm{b}}} = 2.903~{\rm{GeV}},\nonumber\\
\delta_{\rm d}=71.395^{\circ}, ~~\xi_{\rm d}=-0.011473,~~ \zeta_{\rm d}=0.020286\;\nonumber
\end{eqnarray}
leading to $\Delta_{\rm d}=0.01857$. The values for the various CP-angles and  $J_{\rm{CP}}$ corresponding to the above best-fit are observed as
\begin{eqnarray}
\delta_{13}=68.565^{\circ},~~
J_{CP}=2.986\times 10^{-5},\nonumber\\
\alpha=89.538^{\circ},~~\beta=21.932^{\circ},~~\gamma=68.530^{\circ}.\nonumber
\end{eqnarray}
The resulting $V=R^{\dag}O_{\rm d}$ and the corresponding $M^{\prime}_{\rm d}$ (in units of GeV) are
\begin{eqnarray}
\mid V\mid = \left( {\begin{array}{*{20}{c}}
{0.97434}&{0.22507}&0.003553\\
{0.22493}&{0.97350}&{0.04121}\\
0.00863&{0.04045}&{0.99914}
\end{array}} \right), \nonumber ~~
\mid{M^{\prime}_{\rm{d}}}\mid = \left( {\begin{array}{*{20}{c}}
{\bf{0}}&{0.01333}&0.01012\\
{0.01333}&{0.05879}&{0.11728}\\
0.01012&{0.11728}&{2.89816}
\end{array}} \right). \nonumber
\end{eqnarray}

Clearly, these results are in excellent agreement with the predictions in Eqs.(\ref{real md}) and (\ref{predictions}) as well as observed quark masses and mixing data in Eqs. (\ref{PDG best fit}) and (\ref{definition inner angles}). 
\section{Conclusion}
In the absence of a compelling theory of fermion flavor
dynamics from the “top-down” perspective, we have adopted a 'bottom-up' approach to reconstruction of quark mass matrices from physical observables within the framework of Standard Model. The observed data on quark masses and flavor mixing, aided with the standard parametrization for $V$, points to the following model independent and modest texture 4-zero structure of corresponding quark mass matrices in $u$-diagonal basis  i.e. 
\begin{widetext}
\begin{equation}
{M^{\prime}_{\rm{d}}} = \left( {\begin{array}{*{20}{c}}
\bf{0} &{({m_{\rm{s}}}+{m_{\rm{d}}}){s_{12}}}&{{m_{\rm{b}}}{s_{13}}{e^{-i{\delta _{13}}}}}\\
{({m_{\rm{s}}}+{m_{\rm{d}}}){s_{12}}}&{{m_{\rm{b}}}s_{23}^2 + {m_{\rm{s}}}}&{({m_{\rm{b}}}-{m_{\rm{s}}}){s_{23}}}\\
{{m_{\rm{b}}}{s_{13}}{e^{i{\delta _{13}}}}}&{({m_{\rm{b}}}-{m_{\rm{s}}}){s_{23}}}&{m_{\rm{b}}}c_{23}^2
\end{array}} \right)=\left( {\begin{array}{*{20}{c}}
\bf{0}&{{a_{\rm{d}}}}&{{f_{\rm{d}}}{e^{ - i{\delta _{\rm{d}}}}}}\\
{{a_{\rm{d}}}}&{{d_{\rm{d}}}}&{{b_{\rm{d}}}}\\
{{f_{\rm{d}}}{e^{ i{\delta _{\rm{d}}}}}}&{{b_d}}&{{c_{\rm{d}}}}
\end{array}} \right)
\end{equation}
\end{widetext}
which also holds well within the current precision bounds on quark masses and mixing angles. It is observed that certain generalized phenomenological texture zeros may result as a direct consequence of the precision measurements of some physical observables.

Whereas, the Fritzsch-like texture 4-zero Hermitian quark mass matrices involve several parameters and at least two non-trivial phases for consistency with mixing data, the above texture 4-zero structure is quite simpler and entirely expressible in terms of the physical observables. It is thereby shown that the strong hierarchies in the quark mass spectra and mixing angles allows to considerably reduce the 36 arbitrary parameters in the generalized complex mass matrices ${M^{\prime}_{\rm{q}}}$ to only six such parameters namely $a_{\rm{d}}$, $b_{\rm{d}}$, $f_{\rm{d}}$, $d_{\rm{d}}$,  $c_{\rm{d}}$ and $\delta_{\rm{d}}$. Each of these is non-redundant and fixed quite accurately by only six physical observables namely three down sector quark masses, $s_{\rm{23}}$, $s_{\rm{13}}$ and $\delta_{13}$ through uncomplicated yet compelling relations i.e.
\begin{equation}
\begin{array}{c}
a_{\rm{d}}=s_{12}(m_{\rm{d}}+m_{\rm{s}}),\\
b_{\rm{d}}=s_{23}(m_{\rm{b}}-m_{\rm{s}}),\\
f_{\rm{d}}=s_{13} m_{\rm{b}},\\
\delta_{\rm{d}}=\delta_{13},
\end{array}
\end{equation}
and also endorse the GST relation $s_{12}\simeq \sqrt{m_{\rm d}/m_{\rm s}}$ as compared to $s_{12}\simeq \sqrt{m_{\rm d}/m_{\rm s}}-e^{i\phi}\sqrt{m_{\rm u}/m_{\rm c}}$ for Fritzsch-like texture 4-zero matrices. Clearly, the 'up' quark masses are disconnected from flavor mixings and CP-violation in this scenario.

Keeping view of the hierarchy among Yukawa couplings and quark mixing angles and using one-loop renormalization group equations \cite{Babu:1987im, Giudice:1992an} for the Yukawa matrices, the non-leading terms in the Yukawa couplings different from that of the top quark, can be safely neglected. As a result, the running effects on ratio $m_d/m_s$ and hence an emerging texture zero at (11) position in ${M^{\prime}_{\rm{d}}}$ remains invariant to such quantum corrections. 
\begin{acknowledgments}
This work was supported in part by the Department of Science and Technology, India under SERB Research Grant No. SB/FTP/PS-140/2013.
\end{acknowledgments}
\bibliography{thebibliography}
\end{document}